\documentclass[structabstract]{aa}  
\usepackage{graphicx}
\usepackage{txfonts}
\usepackage{natbib}
\usepackage{url}
\usepackage{longtable}
\usepackage{supertabular}

\newcommand{\teff}{T$_{\rm eff}$}
\newcommand{\glog}{log\,g}
\newcommand{\kms}{\,$\mathrm{km\,s^{-1}}$}

\newcommand{\mlp}{\ensuremath{\alpha_{\mathrm{MLT}}}}

\newcommand{\cobold}{{\sf CO$^5$BOLD}}

\chardef\ii="10
\def\i{\'\ii}

\idline{10}{10}

\begin{document}
   \title{Chemical abundances of distant extremely metal-poor 
unevolved stars\thanks{based on spectra obtained with UVES 
at the 8.2m Kueyen ESO telescope, programmes
078.D-0217 and 081.D.0373} }


   \author{P.\,Bonifacio
          \inst{1}
          \and
          L.\,Sbordone\inst{2,1,3}
          \and
          E.\,Caffau\inst{2,1}\fnmsep\thanks{Gliese Fellow}
          \and
          H.-G.\,Ludwig\inst{2,1}
          \and
          M.\,Spite\inst{1}
          \and
          J.\,I.\,Gonz\'alez Hern\'andez\inst{4,5}
          \and
          \hbox{N.\,T.\,Behara\inst{6}}
          }

   \institute{GEPI, Observatoire de Paris, CNRS, Univ. Paris Diderot; Place
Jules Janssen 92190 Meudon, France \\
              \email{Piercarlo.Bonifacio@obspm.fr}
         \and
             Zentrum f\"ur Astronomie der Universit\"at Heidelberg, 
              Landessternwarte, K\"onigstuhl 12, 69117 Heidelberg, Germany
          \and
Max-Planck Institut f\"ur Astrophysik, Karl-Schwarzschild-Str. 1, 85741 Garching, Germany
          \and
Instituto de Astrof\i sica de Canarias (IAC), E-38205
La Laguna, Tenerife, Spain;
          Instituto de Astrofisica de Canarias
\and
Depto. Astrof\i sica, Universidad de La Laguna (ULL),
E-38206 La Laguna, Tenerife, Spain
          \and
Institut d'Astronomie et d'Astrophysique, Universit\'e Libre de Bruxelles, B-1050, Belgium
             }

   \date{Received; accepted }

 
  \abstract
   {The old Galactic halo stars hold the fossil record of  the interstellar 
medium chemical
composition   at the time of their formation. 
Most of the stars studied so far are relatively near to the Sun, this
prompts the study of more distant stars, both to increase the
size of the sample and to search for possible variations
of abundance patterns 
at greater distances.}
   {The purpose of our study 
is to determine the chemical 
composition of a sample of 16 candidate Extremely Metal-Poor (EMP) dwarf
stars, extracted from the Sloan Digital Sky Survey (SDSS). There
are two main purposes: in the first place to verify
the reliability of the metallicity estimates derived from the SDSS spectra;
in the second place to see if the abundance trends found for the
brighter nearer stars studied previously also hold for this sample
of fainter, more distant stars.
}
   {We used the UVES at the VLT to obtain high-resolution spectra
of the programme stars. The abundances were determined by an 
automatic analysis with the {\tt MyGIsFOS \relax}\relax code, with the exception
of lithium, for which the abundances were determined from the measured
equivalent widths of the \ion{Li}{i} resonance doublet. }
   {All candidates are confirmed to be EMP stars, with [Fe/H]$\le -3.0$.
The chemical composition of the sample of stars is similar to that of
brighter and nearer samples.
We measured the lithium abundance for 12 stars and  provide stringent upper
limits for three other stars, for a fourth star the upper limit is not
significant, owing to the low signal-to noise ratio of the spectrum. 
The ``meltdown'' of the Spite plateau is confirmed,
but some of the lowest metallicity stars of the sample lie on the plateau.
}
   {The concordance of the metallicities derived from high-resolution spectra
and those estimated from the SDSS spectra suggests that the latter may
be used to study the metallicity distribution of the halo. 
The abundance pattern  suggests that the halo was well mixed
for all  probed metallicities and distances. The fact that at the
lowest metallicities we find stars on the Spite plateau suggests that 
the meltdown depends on at least another parameter, besides metallicity.}


   \maketitle
%

\section{Introduction}

The Galactic halo contains some of the oldest stars in the Milky
Way, and understanding its formation and evolution
is  one of  the necessary steps to understand the Galaxy as a whole.
Our ideas evolved in time, from the classical ``monolithic collapse''
scenario \citep{ELS}, through the ``chaotic'' scenario
\citep{SZ78} to the realisation that the halo
has had a complex
history and that both collapse  and merging have contributed
to its shaping \citep[for a review see][and references therein]{helmi2008}.

A milestone in the field was certainly the {\em in situ}
study of the north Galactic pole by \citet{Majewski} who showed
that the halo appears to be counter-rotating.
The panoramic view offered by modern wide-field surveys
has allowed us to identify significant sub-structure in the 
halo, often referred to as the ``field of streams'' \citep{belokurov}.
From the theoretical point of view, modern simulations
predict that it is a general expectation that halos of spiral
galaxies are built by a combination of {\em in situ} star formation
and accretion from satellite dwarf galaxies \citep{zolotov,font}.

Exploiting the data of the Sloan Digital Sky Survey 
\citep[SDSS and SEGUE][]{sdss,segue},
\citet{Carollo07,Carollo10} showed that the Galactic stellar
halo is composed of at least two components with distinct kinematics
and metallicity distribution, the more distant halo being
more metal-poor. 
This analysis was based on relatively bright stars in SDSS, for which 
accurate proper motions could be derived.  
The SDSS catalogue  contains spectra and photometry down to $g=20$,
suggesting that an {\em in situ} study of the distant halo is indeed
possible.

With this in mind we began in 2006 to work on the analysis of
SDSS spectra with two main goals: 1) retrieve from SDSS candidate stars
of extremely low metallicity for high-resolution follow-up;
2) determine the metal-weak tail of the halo metallicity   
distribution function directly from the SDSS spectra.

The first goal was motivated to see if there is indeed a metallicity
threshold for  the formation of low-metallicity stars, 
as suggested by some theories of star formation 
at low metallicity \citep{Schneider,BL,Salvadori},
and to check the hypothesis that at very low abundances of iron
only stars with strong enhancements of C and O 
can be found \citep{BL,Frebel}. 

The second goal is straightforward. The metallicity distribution
function (MDF) provides a strong constraint on the theories of 
galaxy formation and evolution, especially at the low-metallicity
end. 
The current versions of the MDF are based on a limited number of
stars \citep{RN91,HES_MDF1,HES_MDF2}, a few thousands at most.
The SDSS has the potentiality of doing this for a sample of stars
that is two orders of magnitudes larger. 
For this to be possible, though, two steps are necessary:
1) the abundance estimated from the low-resolution SDSS
spectra must be confidently calibrated against 
abundances derived from high-resolution spectra; 2) the selection biases
in the SDSS sample must be understood.

It is clear that our goals require that EMP candidates
extracted from the SDSS should be observed at higher resolution.
To this end  we carried out several observational 
campaigns at the European Southern Observatory with
the UVES \citep{uves} and X-Shooter \citep{X-Shooter}
spectrographs. Over the years
we have made three  progress reports
\citep{Ludwig08,bonifacio_rio,sbordone_nic} on this programme. 
The abundances of three C-enhanced dwarf stars, observed with UVES, have been 
reported in \citet{Behara},
and the abundances for two stars observed with X-Shooter  
in \citet{bonifacio11}.
In this paper we report the abundances for 16 stars
of typical magnitude $g\sim 17$,
observed with UVES in the course of ESO periods 78 and 81.
 In an accompanying papers \citep{Elisabetta} we reported
the abundances of five stars observed with X-Shooter. Finally, 
in \citet{EC_Nature,EC12}  
we described the most metal-poor star found 
in the sample, observed with both X-Shooter and UVES.

\section{Target selection}

The programme stars
were selected from the SDSS Data Releases
5 \citep{dr5} and 6 \citep{dr6}. Names, coordinates, 
and $g$ magnitudes
are provided in Table \ref{basic}.
To obtain spectra of good quality in a limited
amount of time with UVES, based on the exposure
time calculator and on our personal experience
with this instrument, we restricted the sample
to $g\le 17.5$. 
We decided to focus on turn-off (TO) stars, for several reasons.
In the first place  
the surface gravity can be confidently fixed at \glog = 4.0
for these stars,
a minor contaminant being Horizontal Branch stars
of higher luminosity.
According to the Padova isochrone of
metallicity --2.0 and age of 13 Gyr \citep{marigo},
the absolute $g$ magnitude of TO stars
ranges between 3.0 and 3.6. This implies, if one 
neglects interstellar extinction, a distance range
between roughly 6 kpc and 8 kpc, these objects are thus
capable to probe the outer halo. 
In this phase of the project we aimed to avoid the
complication of
discriminating between K dwarfs and K giants, 
although in the future we do intend to tackle cooler
stars, because K giants allow one to probe considerably
greater distances.
Another reason is that the SDSS spectroscopic sample
is very rich in TO-colour objects. There are two factors
that contributed to boost their number: metal-poor F dwarfs
were often observed as suitable flux-calibration standards,
and the SDSS QSO selection algorithm \citep{Richards}
also selects very metal-poor TO stars. 

To select the TO stars we relied on the $(g-z)_0$ colour,
which we calibrated against \teff\ \citep{Ludwig08},
our initial selection was $-0.3 \le (g-z)_0 \le 0.7$,
which, in terms of \teff , corresponds to:
$\rm5500 \le$ \teff $\rm\le 8000 K$.
Our initial sample consisted of about 34\,000 stars.
The motivation  for including these high effective
temperatures was to be able to capture the
extremely metal poor stars with C-enhanced 
composition predicted by the isochrones of
\citet{piau}.

   \begin{figure}
   {\centering
   \resizebox{\hsize}{!}{\includegraphics[clip=true]{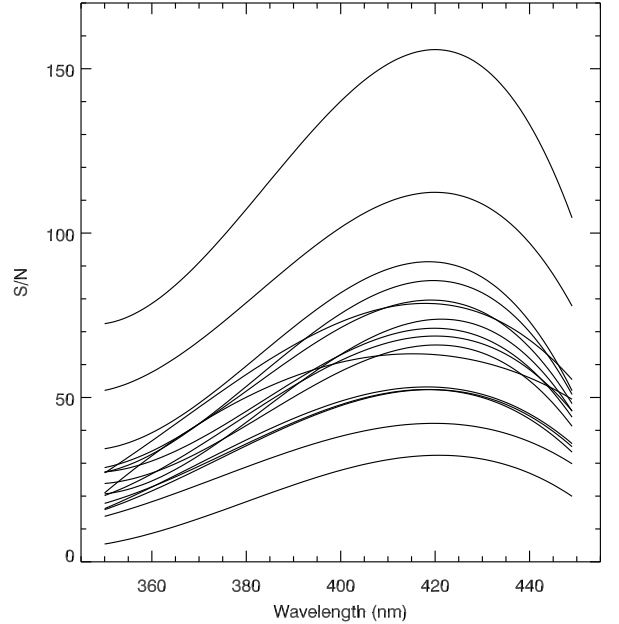}}
}   \caption{The signal-to-noise ratio as a function
of wavelength in the blue-arm spectra. This is a by-product of
{\tt MyGIsFOS}, which calculates it in all the continuum regions. \label{sn}}
\end{figure}

   \begin{figure}
   {\centering
   \resizebox{\hsize}{!}{\includegraphics[clip=true]{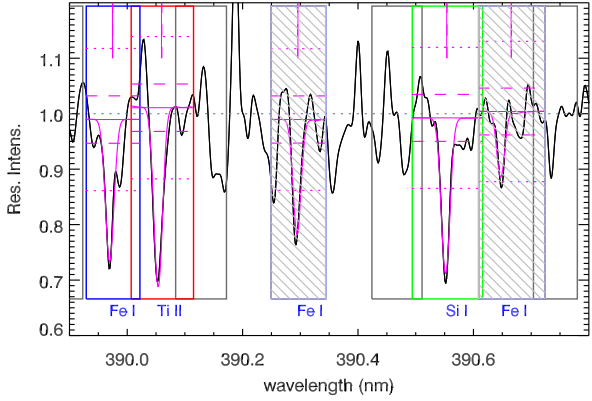}}
}   \caption{ 
Portion of the spectrum of SDSS J1229+2624, one of the lowest
S/N spectra in our sample. The boxes show the wavelength intervals
used to fit the lines  (blue for \ion{Fe}{i} lines, green the \ion{Si}{i} line,
red the \ion{Ti}{ii} line), when they are shaded it means that MyGIsFOS has
rejected the line. Two boxes showing continuum regions are also shown
(grey boxes).
Horizontal dashed and dotted lines indicate the local 1 $\sigma$ and 
3 $\sigma$ estimated
noise ranges. Vertical magenta continuous and dashed lines (in
the top part of the plot) indicate by how much every single line has been
shifted in radial velocity by MyGIsFOS to achieve an optimal fit.
\label{specbad}}
\end{figure}

   \begin{figure}
   {\centering
   \resizebox{\hsize}{!}{\includegraphics[clip=true]{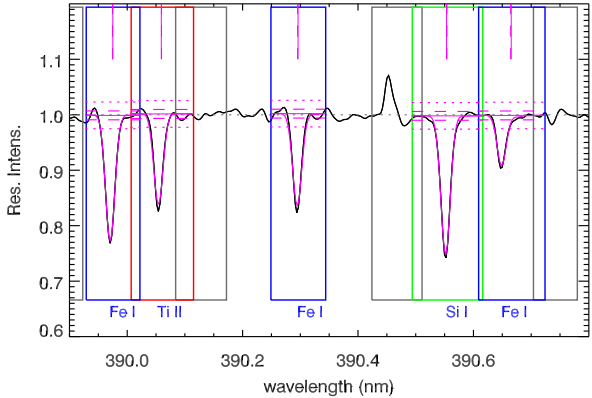}}
}   \caption{Same as Fig.\,\ref{specbad}, but for
\object{SDSS\,J0040+1604} observed in period 81, probably the best
spectrum of the sample. In this case none
of the lines is rejected.  \label{specgood}}
\end{figure}

The raw metallicity distribution function is 
shown in figure 3 of \citet{Ludwig08}.
The almost flat behaviour at metallicities below --4.0 is
clearly suspicious.
It should be emphasized that at these low metallicities
the only metallic line measurable on SDSS spectra
for TO stars is the \ion{Ca}{ii} K line.
Visual inspection of all SDSS spectra of the sample, 
with estimated metallicity below --3.0 showed that in fact 
the low-metallicity sample was dominated by white dwarfs,
some of which do show a measurable  \ion{Ca}{ii} K line in 
the SDSS spectra. At this stage we also realised that
the vast majority of the white dwarfs could indeed
be excluded by a simple cut on the $(u-g)_0$ colour:
$(u-g)_0 >0.70$.
The cleaned sample  with a more restrictive temperature
cut  $0.18 \le (g-z)_0 \le 0.70$ consisted of about 26\,000
stars. The metallicity distribution function was presented
in \citet{bonifacio_rio} and shows no flattening at
low metallicity.
We have already used these cuts on data of SDSS Data Release 7
\citep[see][]{Elisabetta} and we will use it on
SDSS Data Release 8.
These more restricted colour cuts will not be able
to select the stars predicted by \citet{piau}, if they exist.
We point out that finding these objects among the white dwarfs
of similar colours is fairly challenging.

The final selection of the targets to be observed with UVES had to 
take into account the observability conditions 
from Paranal ($\delta \le +28$) and was made by visual inspection
of all stars with [Fe/H]$_{\rm SDSS} \le -3.0$.
It was not our purpose to observe a sample of stars with
a well-defined selection  function, but rather
to observe a sample of true EMP stars.
The three stars reported in \citet{Behara} were already recognized
at this stage as C-enhanced stars. 

For the reader's convenience 
an excerpt of the SDSS photometry for our programme stars is
provided in Table \ref{photo}. Full information is available
on the SDSS site \url{www.sdss.org}.

\section{UVES observations and data reduction}

The observations  were obtained 
in the course of two ESO periods, both programmes
had H.-G. Ludwig as PI. In period 78 we were allocated
29 hours in service mode, in period 81 we
were allocated three nights in visitor mode.
A log of the observations, including the slit width and
CCD binning is given in Table \ref{logobs},
more details on each observation are available through
the ESO archive\footnote{
\url{http://archive.eso.org/wdb/wdb/eso/eso_archive_main/query?prog_id=078.D-0217(A)}
\\
\url{http://archive.eso.org/wdb/wdb/eso/eso_archive_main/query?prog_id=081.D-0373(A)}
\\
\url{http://archive.eso.org/wdb/wdb/eso/eso_archive_main/query?prog_id=081.D-0373(B)}
}
.
We only remark that in the visitor run of April 2008 (observer
L. Sbordone) we lost almost half of the time, due to
strong northern wind, which prohibited pointing of our targets.
The run of August 2008 (observer H.-G. Ludwig) was instead
quite succesful.
The instrumental set-up was similar for the service and the
visitor-mode observations. In both cases
we used the dichroic \# 1 390+580\,nm setting.
In the service mode we used a 1\farcs{4} wide slit
and $2\times 2$ on-chip binning. This was made to 
collect as many photons as possible for these faint stars, 
however, if the seeing was less than the slit width,
the actual resolution was set by the seeing. 
In the visitor observations, the same philosphy 
was employed, but if the observer noticed that 
the seeing was much smaller than the slit width
he had the opportunity of narrowing the slit width
and, if the chosen slit width was less than, or equal to 0\farcs{8},
also to change the binning to $1\times 1$ binning.
The data of period 78 were retrieved reduced from the 
ESO archive, we reduced the observations of period 81
using the UVES pipeline.

For each star we had several spectra
of slightly different resolution. 
To combine the spectra, we estimated the actual resolution
from a number of unblended lines and  then 
convolved each spectrum with a Gaussian so that the 
final resolution was either 13.9 \kms or 12 \kms.

Two stars, 
\object{SDSS J002113--005005} and 
\object{SDSS J004029+160416}
were observed both in period 78 and period 81. 
We decided to analyse  the two sets 
of spectra independently  to obtain a consistency check. 

The signal-to-noise ratios obtained are shown in Fig.\ref{sn}
for the blue arm, where this ratio is lowest.

Star \object{SDSS J153110+095255} was observed in ESO period 81, 
a preliminary analysis confirms that its metallicity is about --3.0,
yet the star is double-lined spectroscopic binary. This star
was excluded from the present sample and will be analysed when more
spectra are available.

\setcounter{table}{1}
\begin{table}
\setlength{\tabcolsep}{4pt}
\caption{Adopted solar abundances.\label{solar}}
\centering
\begin{tabular}{ll}
\hline\hline
Element & log(X/H) +12\\
\hline
Mg & 7.54\\
Si & 7.52\\
Ca & 6.33\\
Sc & 3.10\\
Ti & 4.90\\
Cr & 5.64\\
Mn & 5.37\\
Fe & 7.52\\
Co & 4.92\\
Ni & 6.23\\
Sr & 2.92\\
\hline
\end{tabular}
\end{table}

\begin{table}
\setlength{\tabcolsep}{3pt}
\centering
\caption{Sloan photometry for the programme stars.\label{photo}}
\begin{tabular}{rccccc}
\hline
\hline
Star & $r$ & $(g-z)$ & $(g-z)_0$ & $(u-g) $& $(u-g)_0$ \\
\hline
\object{SDSS J002113--005005} &16.468   &0.28   &0.21   &0.94   &0.90\\
\object{SDSS J002749+140418}  &16.537   &0.62   &0.39   &1.06   &0.93\\
\object{SDSS J004029+160416}  &15.279   &0.36   &0.26   &0.94   &0.88\\
\object{SDSS J031745+002304}  &16.367   &0.74   &0.54   &0.95   &0.83\\
\object{SDSS J082118+181931}  &16.817   &0.36   &0.30   &0.85   &0.81\\
\object{SDSS J082521+040334}  &16.441   &0.45   &0.35   &0.94   &0.89\\
\object{SDSS J090733+024608}  &16.010   &0.50   &0.44   &0.95   &0.91\\
\object{SDSS J113528+010848}  &16.434   &0.44   &0.38   &0.90   &0.87\\
\object{SDSS J122935+262445}  &16.420   &0.31   &0.25   &0.88   &0.84\\
\object{SDSS J130017+263238}  &15.970   &0.30   &0.28   &0.85   &0.84\\
\object{SDSS J143632+091831}  &15.806   &0.37   &0.30   &0.83   &0.79\\
\object{SDSS J144640+124917}  &15.902   &0.41   &0.36   &0.86   &0.83\\
\object{SDSS J154246+054426}  &16.853   &0.52   &0.36   &0.87   &0.78\\
\object{SDSS J223143--094834} &16.519   &0.54   &0.42   &0.86   &0.79\\
\object{SDSS J230814--085526} &16.147   &0.53   &0.44   &0.82   &0.77\\
\object{SDSS J233113--010933} &17.192   &0.42   &0.34   &0.90   &0.85\\
\hline
\end{tabular}
\end{table}

\begin{table*}
\setlength{\tabcolsep}{3pt}
\centering
\caption{Coordinates and atmospheric parameters for the programme stars.\label{basic}}
\begin{tabular}{rlccccrrrcr}
\hline\hline
Star & $g$ & $\alpha$ & $\delta$ & \teff & log g & $\xi$ & [Fe/H] &  S/N & ESO period\\
     & mag & J2000 & J2000 & K & c.g.s.          & \kms  &                    &    &\multicolumn{2}{r}{SDSS Object Type$^a$}   \\
     &     &       &       &   &                 &       &                        &@420\,nm   \\
\hline

\object{SDSS J002113-005005} &  16.7 & 00:21:13.78 &--00:50:05.2&6546 &4.59& 1.54 &$ -3.15 $   &52 & 78 & QSO\\
                    &      &             &             &6546 &4.01& 1.76 &$ -3.25 $   &74 & 81 &  QSO\\
\object{SDSS J002749+140418}  & 16.9 & 00:27:49.46 & +14:04:18.1&6125 &3.61& 1.46 &$ -3.37 $   &69 & 78& SP STD\\
\object{SDSS J004029+160416}  & 15.5 & 00:40:29.17 & +16:04:16.2&6422 &3.88& 1.48 &$ -3.27 $   &112& 78& Ser BLUE\\
                    &      &             &             &6422 &3.93& 1.58 &$ -3.25 $   &156& 81& Ser BLUE\\
\object{SDSS J031745+002304}  & 16.8 & 03:17:45.82 & +00:23:04.2&5786 &4.02& 1.41 &$ -3.46 $   &66 & 78& SP STD\\
\object{SDSS J082118+181931}  & 16.7 & 08:21:18.18 & +18:19:31.8&6158 &4.00& 1.50 & $-3.80 $   &19 & 78& SP STD\\
\object{SDSS J082521+040334}  & 17.1 & 08:25:21.29 & +04:03:34.4&6340 &4.00& 1.23 &$ -3.46 $   &80 & 81& SP STD\\
\object{SDSS J090733+024608}  & 16.3 & 09:07:33.28 & +02:46:08.2&5934 &3.71& 1.61 &$ -3.44 $   &204 & 78 &SP STD\\ 
\object{SDSS J113528+010848}  & 16.7 & 11:35:28.08 & +01:08:48.8&6132 &3.83& 1.65 &$ -3.03 $   &63 & 78&SP STD \\
\object{SDSS J122935+262445}  & 16.7 & 12:29:35.95 & +26 24 45.9&6452 &4.20& 2.68 &$ -3.29 $   &32 & 81&SP STD\\
\object{SDSS J130017+263238}  & 16.2 & 13:00:17.20 & +26:32:38.6&6393 &4.00& 1.36 &$ -3.65 $   &85 & 81&SP STD\\
\object{SDSS J143632+091831}  & 16.1 & 14:36:32.27 & +09:18:31.5&6340 &4.00& 1.42 &$ -3.40 $   &53 & 78&SP STD \\
\object{SDSS J144640+124917}  & 16.1 & 14:46:40.63 & +12:49:17.5&6189 &3.90& 1.77 &$ -3.16 $   &52 & 78&SP STD\\
\object{SDSS J154246+054426}  & 17.2 & 15:42:46.87 & +05:44:26.4&6179 &4.00& 1.68 &$ -3.48 $   &78 & 78&SP STD\\ 
\object{SDSS J223143--094834} & 16.9 & 22:31:43.95 & -09:48:34.4&6053 &4.16& 1.36 &$ -3.20 $   &91 & 78&SP STD\\
\object{SDSS J230814--085526} & 16.5 & 23:08:14.85 & -08:55:26.4&6015 &4.66& 1.74 &$ -3.01 $   &42 & 78&SP STD\\
\object{SDSS J233113--010933} & 17.5 & 23:31:13.50 & -01:09:33.4&6246 &4.21& 1.38 &$ -3.08 $   &71 & 81&RED STD\\ 
\hline
\\
\multicolumn{11}{l}{$^a$ Object type, based on colours that motivated the selection of this
object for SDSS spectroscopy: QSO = quasi stellar object;} \\
\multicolumn{11}{l}{SP STD = spectrophotometric standard;
Ser BLUE = serendipity blue object; RED STD = reddening standard.}
\end{tabular}
\end{table*}

\section{Abundance analysis}

The analysis was performed with the {\tt MyGIsFOS} code
\citep{sbordone_nic,mygisfos}.
The code  evolved from the code of \citet{BC03}, and was entirely
re-coded in {\tt fortran 90} with several improvements.
Briefly: this program compares the observed spectrum to a grid
of synthetic spectra and performs  $\chi^2$ fitting.
The most noticeable improvement is a loop to determine the surface gravity
from the iron ionisation equilibrium.
The grid of synthetic spectra was computed from a grid
of ATLAS 12 
\citep{K05,Castelli05,SB04,SB05} 
model atmospheres.
Convection was treated in the mixing-length
approximation with \mlp = 1.25.
The grid covers the effective temperatures 5400 K -- 7000 K
at steps of 200 K, logarithmic  surface gravities  3.5 -- 4.5
(c.g.s units) at steps of 0.5\,dex, logaritmic metallicities n the range 
--4.0  to --2.0, at steps of 0.5\,dex,
[$\alpha$/Fe]=0.0,+0.4 and microturbulent velocities of 1,2,3 \kms.

The effective temperature was considered a prior and was 
derived from the $(g-z)_0$ colour using the calibration
presented in \citet{Ludwig08}. 
We checked the effective temperatures also with fits
of the wings of H$\alpha$ based on \cobold\ models, 
as in \citet{sbordone10}, these provide
temperature that are hotter, on average, by 116\,K.
In five cases the fits required extrapolation
beyond the range of the CIFIST grid \citep{Ludwig09}.
We decided to keep the photometric temperatures, 
to make the comparison
of the metallicities with those derived from the
SDSS spectra straightforward.
The surface gravity was determined
from the iron ionisation equilibrium. If no \ion{Fe}{ii} lines
were retained in the analysis, then the surface gravity was
held fixed at the starting value, \glog = 4.0.
The ratio [$\alpha$/Fe] was determined from the Mg and Ca lines.
The solar abundances adopted are given in Table\,\ref{solar},
Fe is from \citet{caffau_sol}, the other elements
are from \citet{lodders}.

Two examples of fits performed by  {\tt MyGIsFOS} 
are shown in figures \ref{specbad} and \ref{specgood},
illustrating the capability of {\tt MyGIsFOS} to reject
poor fits. 
That we recovered the same abundances,
within errors, for the two stars for which 
we analysed the spectra observed in period
78 and 81 independently,  reassures us about  
the robustness and reproducibility
of  {\tt MyGIsFOS}. 

\begin{figure}
   \centering
   \includegraphics[clip=true]{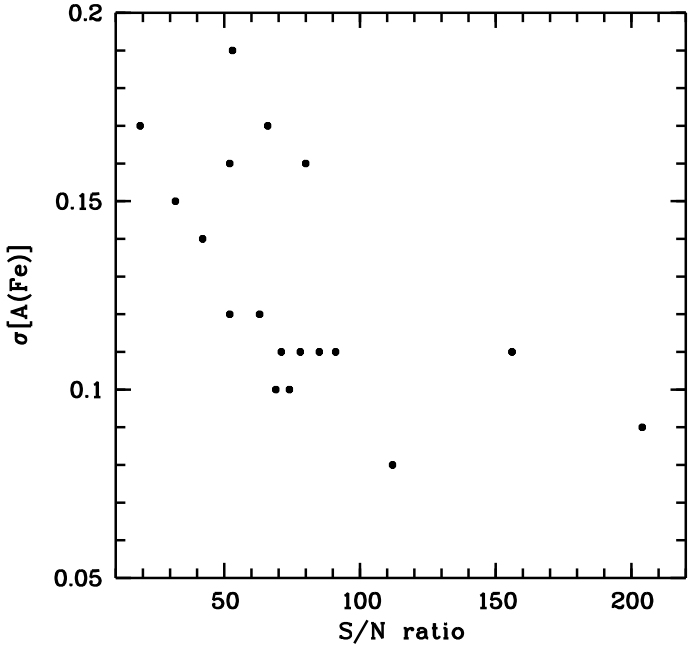}
   \caption{Line-to-line scatter in the \ion{Fe}{i} abundances as 
a function of the S/N ratio at 420\,nm. There is a clear correlation
between the two.}
              \label{sigsn}%
    \end{figure}

The resulting chemical abundances are provided in Tables \ref{alpha}
and \ref{ironpeak}. For chemical species for which several lines
are measurable, we provide the standard deviation of the abundances
derived from the different lines (line-to-line scatter). 
This can be assumed to be a reasonable
estimate of the error on the provided abundance or abundance ratio. 
The different quality (S/N ratio) of the spectra is immediately
reflected in the line-to-line scatter of a species with many lines
available, such as \ion{Fe}{i}. This is 
indeed illustrated in Fig. \ref{sigsn}. 
For species for which only one line is available the error could
be estimated from the S/N in the spectrum at the wavelength of the line. 
With the spectra at hand this should be in the range 0.1-0.2\,dex.

The  systematic error caused by the uncertainty in atmospheric
parameters has been treated in many papers. For an estimate
we refer the readers to Table 4 of \citet{bonifacio09},
the systematic errors in our analysis are essentially the same.

For lithium we did not use
{\tt MyGIsFOS}, but we
measured the equivalent widths by fitting synthetic profiles, 
as described in \citet{bonifacio02} and then determined 
the lithium abundances by using the fitting formula of
\citet{sbordone10}, based on \cobold\ 3D models
\citep{freytag02,wedemeyer04,freytag11} and NLTE treatment
of line transfer.
For the stars for which we could not detect the Li doublet
we estimated the upper limit at 3$\sigma$ by
using the Cayrel formula \citep{cayrel88}. 
The results are provided in Table \ref{lithium}.

Star \object{SDSS J090733+024608}  is in common
between us and \citet{Elisabetta}, 
the abundances derived by us from the UVES spectrum of this star agree 
excellently with those derived by
\citet{Elisabetta} from the X-Shooter spectrum.

   \begin{figure}
   {\centering
   \includegraphics[clip=true]{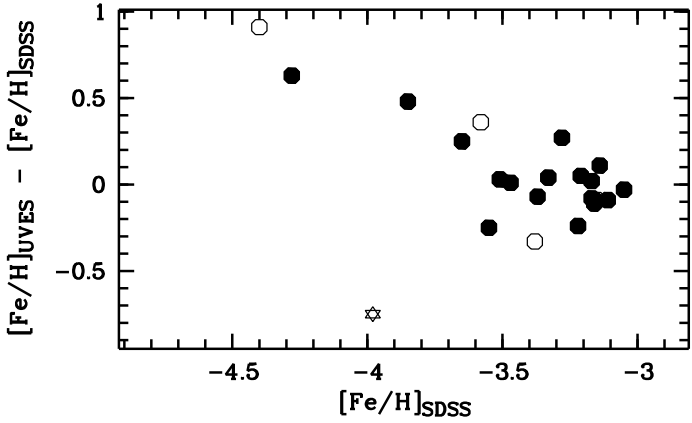}
}   \caption{Difference between the
[Fe/H] derived from the UVES spectra and that derived
from the SDSS spectra as a function of \hbox{[Fe/H]$_{\rm SDSS}$}.}
Eviodently down to [Fe/H] $\sim -3.5$
is accurate to about 0.2\, dex, but rapidly increases
below. The open symbols are the stars observed
with X-Shooter by \citet{Elisabetta}.
The star symbol is SDSS\,J102915+172927 
\citep{EC_Nature}.
\label{figmet}
\end{figure}

   \begin{figure}
   \centering
   \includegraphics[clip=true]{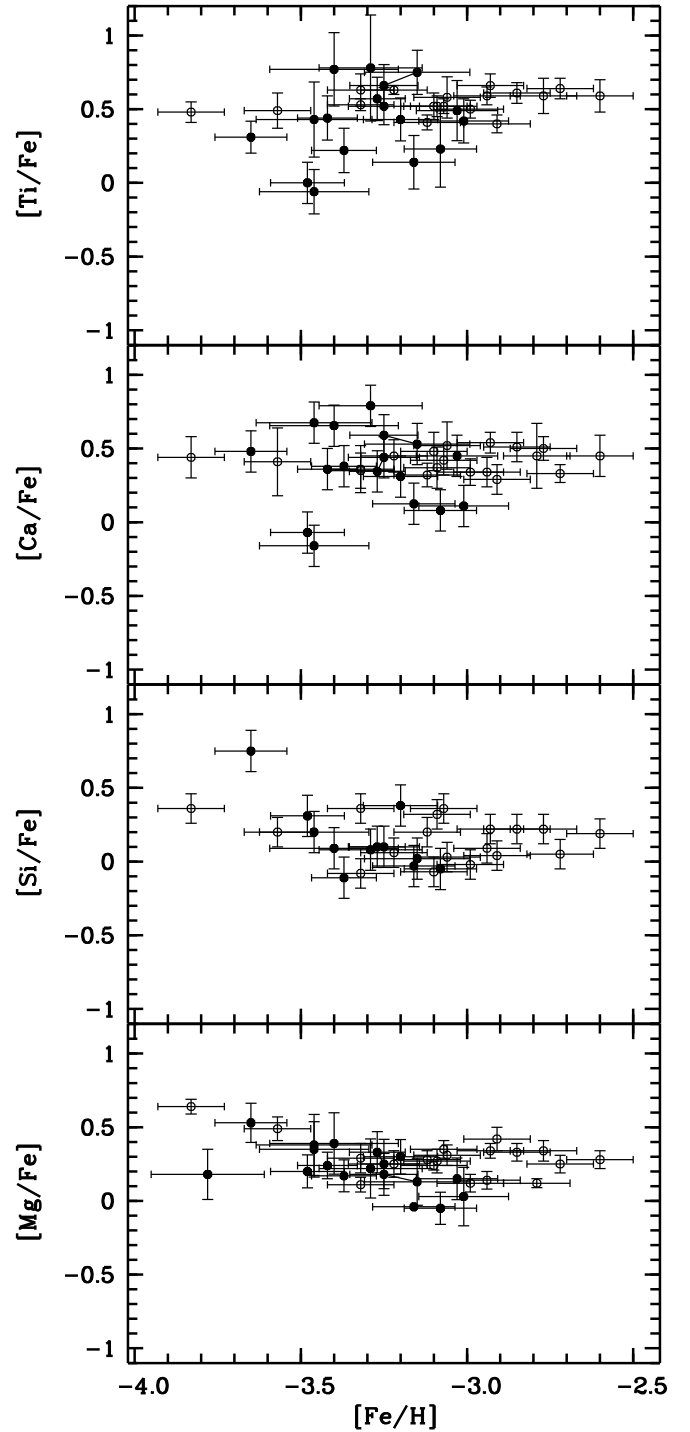}
   \caption{Ratios of alpha elemements to iron for our
programme stars (filled circles) compared to those of the
stars of \citet{bonifacio09} (open circles) The  spectra of
\object{SDSS J004029+160416} and \object{SDSS J002113-005005} observed
in periods 78 and 81 were treated independently
and the points referring to the derived abundances are
connected by a line. The one referring to the two
measures of \object{SDSS J004029+160416} is not visible, since the
two points are indistinguishable at the resolution of the plot.}
              \label{figalpha}%
    \end{figure}

   \begin{figure}
   \centering
   \includegraphics[clip=true]{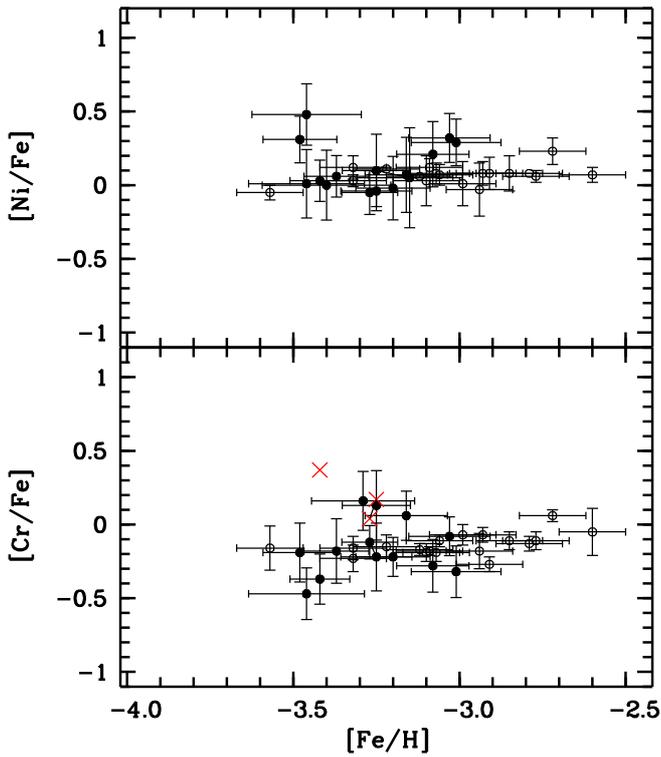}
   \caption{Ratios of even iron-peak elemements to iron for our
programme stars (filled circles) compared to those of the
stars of \citet{bonifacio09} (open circles). 
In the lower panel the two  measures of \ion{Cr}{ii} lines are shown as
$\times$ symbols.}
              \label{fig_iron_even}%
    \end{figure}

   \begin{figure}
   \centering
   \includegraphics[clip=true]{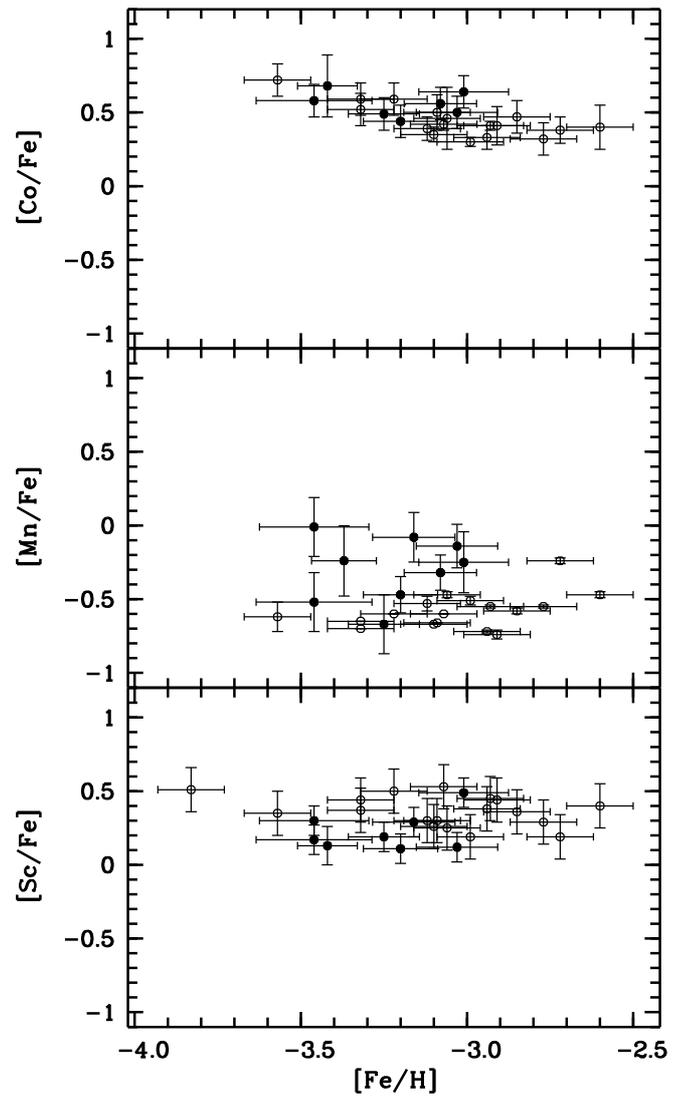}
   \caption{Ratios of odd iron-peak elemements to iron for our
programme stars (filled circles) compared to those of the
stars of \citet{bonifacio09} (open circles). 
}
              \label{fig_iron_odd}%
    \end{figure}

   \begin{figure}
   \centering
   \includegraphics[clip=true]{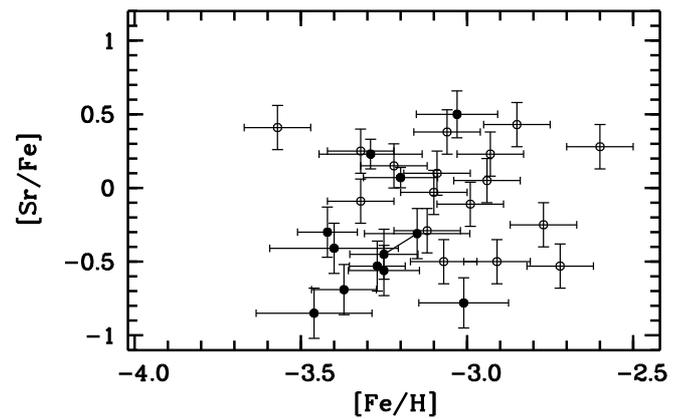}
   \caption{Ratios of strontium to iron for our
programme stars (filled circles) compared to those of the
stars of \citet{bonifacio09} (open circles). 
}
              \label{fig_sr}%
    \end{figure}

   \begin{figure}
   {\centering
   \resizebox{\hsize}{!}{\includegraphics[clip=true]{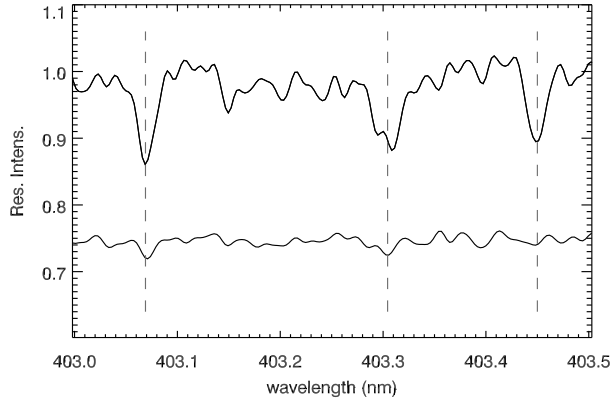}}
}   \caption{
\ion{Mn}{i} 403\,nm resonance triplet plotted for star \object{SDSS\,J144640+124917} (above) and 
\object{SDSS\,J004029+160416}.
The spectra are normalized, and the continuum for star \object{SDSS\,J004029+160416}  
was shifted down to 0.75 for readibility. 
The wavelengths of the three Mn lines are
marked. Star \object{SDSS\,J004029+160416} is 280 K warmer and about 
0.1 dex more metal-poor.
Gravity is almost the same. [Mn/Fe] = -0.08 \relax
in \object{SDSS\,J144640+124917} and -0.67 \relax in \object{SDSS\,J004029+160416}. 
 \label{mnplot}}
\end{figure}

   \begin{figure}
   {\centering
   \resizebox{\hsize}{!}{\includegraphics[clip=true]{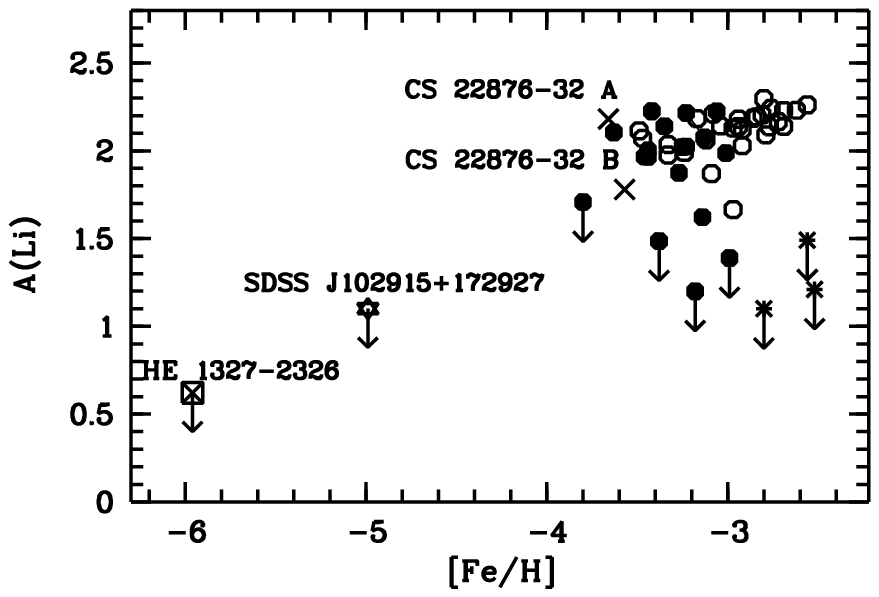}}
}   \caption{Lithium abundance as a function of [Fe/H]
for our sample of stars (filled hexagons), together with the
stars of \citet[][open hexagons]{sbordone10},
the two components of the binary system
\object{CS\,22876-32} \citep[][crosses]{jonay}, the 
three Li-depleted stars \object{G\,122-69},
\object{G\,139-8}, \object{G\,186-26} from \citet[][asterisks]{norris},
star \object{SDSS\,J102915+172927}, from 
\citet[][star symbol]{EC_Nature} and star
\object{HE\,1327-2326} from \citet[][crossed square]{Frebel08}.  \label{limet}}
\end{figure}

   \begin{figure}
   {\centering
   \resizebox{\hsize}{!}{\includegraphics[clip=true]{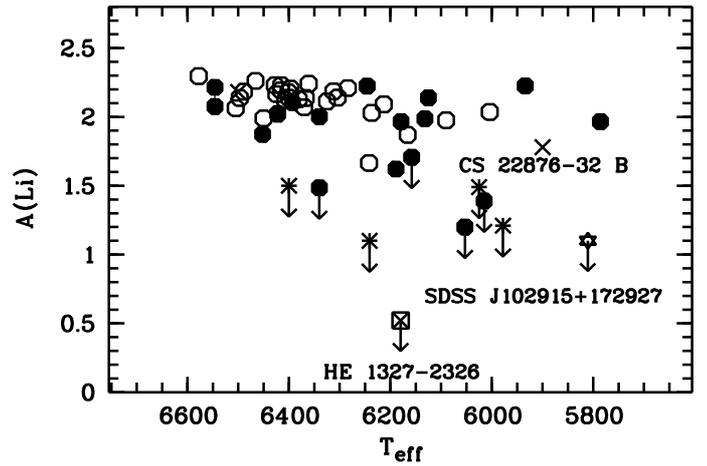}}
}   \caption{Lithium abudance as a function of \teff ,
the symbols are the same as in Fig.\, \ref{limet} 
 \label{liteff}}
\end{figure}

\begin{table*}
\setlength{\tabcolsep}{1pt}
\caption{Abundances of the $\alpha$ elements and strontium.\label{alpha}}
\centering
\begin{tabular}{lcccccccccccccccccccr}
\hline\hline
Star        &[Fe/H]& [Mg/Fe] & $\sigma$ & N & [Si/Fe] & N & [\ion{Ca}{i}/Fe] & $\sigma$ & N & [\ion{Ca}{ii}/Fe] & N &
  [\ion{Ti}{i}/Fe] & $\sigma$ & N & [\ion{Ti}{ii}/Fe] & $\sigma$ & N & [Sr/Fe] & $\sigma$ & N \\ 
\hline
\object{SDSS J002113--005005}  &$ -3.15$&$ +0.13$ &0.16 &3&$+0.02$&1&      &     & &$ +0.53$&1&     &     &  &$+0.75$ &    & 6&$-0.31$&     &2\\ 
                     &$ -3.25$&$ +0.18$ &0.14 &2&$     $& &+0.72 &     &1&$ +0.46$&1&     &     &  &$+0.66$ &0.14& 8&$-0.45$&     &1\\ 
\object{SDSS J002749+140418}  &$ -3.37$&$ +0.17$ &0.11 &3&$-0.11$&1&+0.40 &0.22 &2&$ +0.36$&1&     &     &  &$+0.22$ &    & 6&$-0.69$&     &2\\ 
\object{SDSS J004029+160416}  &$ -3.27$&$ +0.33$ &0.14 &3&$+0.10$&1&+0.28 &0.08 &2&$ +0.41$&1&     &     &  &$+0.57$ &0.15&12&$-0.53$&     &1\\ 
                     &$ -3.25$&$ +0.25$ &0.17 &2&$+0.10$&1&+0.38 &0.11 &2&$ +0.50$&1&+1.09&     & 1&$+0.52$ &0.13&12&$-0.56$&     &1\\ 
\object{SDSS J031745+002304}  &$ -3.46$&$ +0.38$ &0.21 &4&$     $& &+0.60 &     &1&$ +0.75$&1&+1.04&     & 1&$+0.43$ &0.25& 3&$-0.85$&     &1\\ 
\object{SDSS J082118+181931}  &$ -3.80$&$ +0.24$ &0.17 &2            \\  
\object{SDSS J082521+040334}  &$ -3.46$&$ +0.35$ &0.19 &3&$+0.20$&1&      &     & &$ -0.16$&1&     &     &  &$-0.06$ &    & 1&$     $&     & \\ 
\object{SDSS J090733+024608}  &$ -3.44$&$ +0.30$ &0.09 &2&$     $& &+0.51 &     &1&$ +0.31$&1&$+0.82$ & & 1& $+0.58$ &0.15& 11& $-0.23$& & 1\\ 
\object{SDSS J113528+010848}  &$ -3.03$&$ +0.15$ &0.14 &3&$     $& &+0.45 &0.13 &2&$      $& &     &     &  &$+0.49$ &0.20& 9&$+0.50$&0.16 &2\\ 
\object{SDSS J122935+262445}  &$ -3.29$&$ +0.22$ &0.20 &4&$+0.08$&1&      &     & &$ +0.79$&1&     &     &  &$+0.78$ &0.36& 4&$+0.23$&0.10 &2\\ 
\object{SDSS J130017+263238}  &$ -3.65$&$ +0.53$ &0.13 &2&$+0.75$&1&      &     & &$ +0.48$&1&     &     &  &$+0.31$ &0.11& 2&$     $&     & \\ 
\object{SDSS J143632+091831}  &$ -3.40$&$ +0.39$ &0.21 &2&$+0.09$&1&+0.87 &     &1&$ +0.44$&1&     &     &  &$+0.77$ &0.25& 2&$-0.41$&     &1\\ 
\object{SDSS J144640+124917}  &$ -3.16$&$ -0.04$ &     &1&$-0.03$&1&+0.51 &     &1&$ -0.26$&1&     &     &  &$+0.14$ &0.18& 5&$     $&     & \\ 
\object{SDSS J154246+054426}  &$ -3.48$&$ +0.20$ &0.11 &3&$+0.31$&1&      &     & &$ -0.07$&1&     &     &  &$-0.00$ &0.14& 2&$     $&     & \\ 
\object{SDSS J223143--094834} &$ -3.20$&$ +0.30$ &0.12 &2&$+0.38$&1&+0.47 &0.16 &6&$ +0.15$&1&+0.77& 0.18& 2&$+0.43$ &0.15& 6&$+0.07$&0.07 &2\\ 
\object{SDSS J230814--085526} &$ -3.01$&$ +0.03$ &0.20 &3&$     $& &+0.28 &0.14 &2&$ -0.06$&1&     &     &  &$+0.42$ &    & 7&$-0.78$&     &1\\ 
\object{SDSS J233113--010933} &$ -3.08$&$ -0.05$ &0.11 &2&$-0.05$&1&      &     & &$ +0.08$&1&+0.93&     & 1&$+0.23$ &0.26& 5&$     $&     & \\ 
\hline
\end{tabular}
\end{table*}

\begin{table*}
\setlength{\tabcolsep}{1pt}
\caption{Abundances of the iron peak elements.\label{ironpeak}}
\centering
\begin{tabular}{lcccccccccccccccccccr}
\hline\hline
Star      & [\ion{Fe}{i}/H] & $\sigma$ & N & [\ion{Fe}{ii}/H] & $\sigma$ & N & [Sc/Fe] & N & [Cr/Fe] & $\sigma$ & N
& [Mn/Fe]& $\sigma$ & N & [Co/Fe] & $\sigma$ & N & [Ni/Fe] &$\sigma$ & N \\
\hline
\object{SDSS J002113--005005} &$-3.15$  &0.16 &36&$-3.15$ &     &1&$     $& &$     $&    &   &$     $&     & &$     $&     & &$+0.05$&0.34& 5\\
                    &$-3.25$  &0.10 &38&$-3.26$ &0.12 &3&$     $& &$+0.13$&0.24&  2&$     $&     & &$     $&     & &$+0.10$&0.25& 4\\
\object{SDSS J002749+140418} &$-3.37$  &0.10 &40&$-3.37$ &     &1&$     $& &$-0.18$&0.22&  4&$-0.24$&0.22 &2&$     $&     & &$+0.06$&0.14& 5\\
\object{SDSS J004029+160416} &$-3.27$  &0.08 &43&$-3.27$ &0.09 &3&$     $& &$-0.12$&0.11&  4&$     $&     & &$     $&     & &$-0.05$&0.15& 6 \\ 
                    &$-3.25$  &0.11 &43&$-3.25$ &0.05 &2&$+0.19$&1&$-0.22$&0.23&  3&$-0.67$&     &1&$+0.49$&     &1&$-0.04$&0.13& 8 \\
\object{SDSS J031745+002304} &$-3.46$  &0.17 &42&$-3.46$ &0.13 &3&$+0.17$&1&$-0.47$&0.18&  2&$-0.52$&     &1&$+0.58$&     &1&$+0.01$&0.23& 8 \\
\object{SDSS J082118+181931} &$-3.80$  &0.17 &5 & \\ 
\object{SDSS J082521+040334} &$-3.46$  &0.16 &36&$     $ &     & &$+0.30$&1&$     $&    &   &$-0.01$&     &1&$     $&     & &$+0.48$&0.21& 4\\
\object{SDSS J090733+024608} &$-3.44$  &0.09 &47&$ -3.44$ &    &1&$+0.22$&3 &$ -0.32$ & 0.17 &5& & && +0.70  & 0.21 & 3& +0.07 & 0.14 & 7\\    
\object{SDSS J113528+010848} &$-3.03$  &0.12 &54&$-3.03$ &0.11 &2&$+0.12$&1&$-0.08$&0.13&  3&$-0.14$&0.08 &2&$+0.50$&     &1&$+0.32$&0.17& 9 \\
\object{SDSS J122935+262445} &$-3.29$  &0.15 &29&$-3.29$ &0.03 &2&$     $& &$+0.16$&    &  1&$     $&     & &$     $&     & &$     $&    &  \\
\object{SDSS J130017+263238} &$-3.65$  &0.11 &30&$     $ &     & &$     $& &$     $&    &   &$     $&     & &$     $&     & &$     $&    &  \\
\object{SDSS J143632+091831} &$-3.40$  &0.19 &34&$     $ &     & &$     $& &$     $&    &   &$     $&     & &$     $&     & &$+0.00$&0.24& 2\\
\object{SDSS J144640+124917} &$-3.16$  &0.12 &40&$-3.15$ &0.06 &2&$+0.29$&1&$+0.06$&0.17&  3&$-0.08$&0.11 &3&$     $&     & &$+0.07$&0.25& 6\\
\object{SDSS J154246+054426} &$-3.48$  &0.11 &34&$     $ &     & &$     $& &$-0.19$&    &  1&$     $&     & &$     $&     & &$+0.31$&0.16& 6\\
\object{SDSS J223143--094834}&$-3.20$  &0.11 &47&$-3.20$ &0.03 &3&$+0.11$&1&$-0.22$&0.13&  4&$-0.47$&0.05 &2&$+0.44$& 0.12&3&$-0.02$&0.22&10 \\
\object{SDSS J230814--085526}&$-3.01$  &0.14 &50&$-3.01$ &     &1&$+0.49$&1&$-0.32$&0.17&  3&$-0.25$&0.16 &2&$+0.64$&     &1&$+0.29$&0.16& 7\\
\object{SDSS J233113--010933}&$-3.08$  &0.11 &42&$-3.08$ &0.24 &2&$     $& &$-0.28$&0.18&  4&$-0.32$&0.05 &3&$+0.56$&     &1&$+0.21$&0.22& 6\\
\hline
\end{tabular}
\end{table*}

\begin{table}
\setlength{\tabcolsep}{4pt}
\caption{Abundances of lithium  in the programme stars.\label{lithium}}
\centering
\begin{tabular}{lrr}
\hline\hline
 Star                &   EW (pm) & A(Li) \\           
\hline
\object{SDSS J002113--005005}  &  $ 1.26$ &$ 2.08$\\ 
                     &  $ 1.69$ &$ 2.21$\\
\object{SDSS J002749+140418}  &  $ 2.60$ &$ 2.13$\\
\object{SDSS J004029+160416}  &  $ 1.31$ &$ 2.02$\\
                     &  $ 1.32$ &$ 2.03$\\
\object{SDSS J031745+002304}  &  $ 3.05$ &$ 1.97$\\
\object{SDSS J082118+181931}  &  $<0.98$ &$ <1.71$\\
\object{SDSS J082521+040334}  &  $ 1.43$ &$ 2.02$\\
\object{SDSS J090733+024608}  &  $ 4.15$ &$ 2.23$\\
\object{SDSS J113528+010848}  &  $ 1.87$ &$ 1.99$\\
\object{SDSS J122935+262445}  &  $ 0.92$ &$ 1.88$\\
\object{SDSS J130017+263238}  &  $ 1.66$ &$ 2.10$\\
\object{SDSS J143632+091831}  &  $<0.45$ &$<1.48$\\
\object{SDSS J144640+124917}  &  $ 0.77$ &$ 1.62$\\
\object{SDSS J154246+054426}  &  $ 1.68$ &$ 1.97$\\
\object{SDSS J223143--094834} &  $<0.37$ &$<1.20$\\
\object{SDSS J230814--085526} &  $<0.60$ &$<1.39$\\
\object{SDSS J233113--010933} &  $ 2.64$ &$ 2.22$\\
\hline
\end{tabular}
\end{table}

\section{Distances}

Using the SDSS photometry and theoretical isochrones,
it is possible to estimate the distances of our sample of stars.
We used the Padova isochrone \citep{girardi} with $Z=0.0001$
and an age of 13.5 Gyr.
To read the absolute magnitude from the isochrone it
is necessary to decide whether a star is above
or below the turn-off. For any set of atmospheric
parameters we assumed the star is on the Main Sequence (MS, 
i.e. it is {\em below} the TO) if \glog $\ge 4.18$
and on the sub giant branch (SG, i.e. is {\em above} the TO) if
\glog $< 4.18$. This distinction is based
on the estimated surface gravity of the TO from
the theoretical isochrone.
In Table \ref{distance} we provide the
distances, $d$, from the Sun derived in this way, averaged
over the five SDSS bands, in kpc.
In the column $\sigma_d$ we provide the standard
deviation of the distances derived from the different
bands.
This should not be used as an estimate for the
errors on our distances, but only as a sanity
check of the consistency of the photometry in the 
different bands.
For star 
\object{SDSS J090733+024608} \citet{Elisabetta}
estimate a distance that is about 1 kpc larger, becasue they chose 
a different  isochrone. This difference
can be taken as an estimate of the uncertainty on our distances.

We also provide the distances from 
the Galactic centre, computed as 
$R = \sqrt { d^{\,2} - 2 d R_{GC}\cos(b)\cos(l) + R_{GC}^{\,2} }$
where $R_{GC}$ is the Sun-Galactic centre  distance,
taken to be 8.5 Kpc, $l,b $  are Galactic longitude and latitude.
As expected from their faintness, the stars are distant. 

In the last column of Table \ref{distance} we also provide
the barycentric radial velocities, computed by
cross-correlation of the blue spectra against
a synthetic template, after masking out the Balmer lines.
For the stars for which we have multiple observations
the standard deviation of the different radial velocities
never exceeds 0.6 \kms , thus none of the stars shows
any radial velocity variations. 
The radial velocity accuracy is dominated by the systematics, 
and mainly by the centering of the star in the slit.
The plate scale of the UVES blue arm (0\farcs{215} or 0.0019 nm in 
the dispersion direction) implies that an error of 0\farcs{2} in centering,
corresponds to 1.3 \kms, this can be taken as an estimate
of the absolute error on  radial velocities.
Our measured radial velocity for \object{SDSS J090733+024608} 
agrees excellently with the measurement of \citet{Elisabetta}
from their X-Shooter spectrum.

\citet{bonifacio09} did not estimate distances for their sample of stars, 
and it is beyond the purpose of the present paper to provide such
estimates for those stars. Suffice it to say that the V magnitudes
of the \citet{bonifacio09} sample range from 13 to 15.2, while most
of the stars in the present sample have $g$ magnitudes between 16
and 17.5. Given that $V$ and $g$ magnitudes are fairly similar for
stars of these spectral types, we expect the stars of the present
sample to be on average four times more distant\footnote{If two stars have
the same absolute magnitude and different apparent magnitudes $m_1$ and
$m_2$, the ratios of their distances is $\log (d_1/d_2) = (m_1-m_2)/5$. 
Thus if $m_1-m_2 =3 $, then  $d_1 \approx 4 d_2$}.

\begin{table*}
\setlength{\tabcolsep}{3pt}
\centering
\caption{Galactic coordinates, distances from the Sun and from the
Galactic centre, 
and barycentric radial velocities,  for the programme stars.\label{distance}}
\begin{tabular}{rrrcrrrr}
\hline\hline
Star & $l$ & $b$ & Class& $d$    & $\sigma_d$ & $R$   & $v_R$  \\
     &      &     &      & Kpc  & Kpc        &  Kpc& \kms    \\
\hline

\object{SDSS J002113-005005}   & 106.2675&--62.7251& MS & 3.96 & 0.07 & 9.82   & --94.2 \\ 
                               & 106.2675&--62.7251& SG & 2.47 & 0.06 & 9.15   &        \\
\object{SDSS J002749+140418}   & 114.2890&--48.4034& SG & 4.30 & 0.07 & 10.52  & +27.8  \\
\object{SDSS J004029+160416}   & 119.0930&--46.7189& SG & 2.32 & 0.04 & 9.53   & --49.1 \\
                               & 119.0930&--46.7189& SG & 2.32 & 0.04 & 9.53   &        \\
\object{SDSS J031745+002304}   & 180.9504&--45.3647& SG & 4.29 & 0.07 & 11.91  & +113.5 \\
\object{SDSS J082118+181931}   & 205.5011& +27.7892& SG & 4.41 & 0.05 & 12.31  & +162.6 \\
\object{SDSS J082521+040334}   & 220.2941& +22.7044& SG & 5.00 & 0.03 & 12.53  & +10.8  \\
\object{SDSS J090733+024608}   & 227.2893& +31.2990& SG & 3.82 & 0.03 & 11.16  & +313.0 \\ 
\object{SDSS J113528+010848}   & 264.7169& +58.2657& SG & 4.46 & 0.03 & 9.79   & --88.9 \\
\object{SDSS J122935+262445}   & 220.0458& +85.0737& MS & 2.20 & 0.03 & 8.92   & +66.3  \\
\object{SDSS J130017+263238}   &  16.9469& +87.9415& SG & 3.44 & 0.02 & 9.06   & --77.4 \\
\object{SDSS J143632+091831}   &   2.2179& +59.4609& SG & 3.13 & 0.02 & 7.42   & --113.2 \\
\object{SDSS J144640+124917}   &  10.6982& +59.4938& SG & 3.45 & 0.02 & 7.41   & --110.4 \\
\object{SDSS J154246+054426}   &  13.1225& +44.0248& SG & 5.08 & 0.05 & 6.13   & --123.9 \\ 
\object{SDSS J223143--094834}  &  53.9497&--52.8171& SG & 4.59 & 0.05 & 8.10   & --5.6  \\
\object{SDSS J230814--085526}  &  64.8229&--59.5785& MS & 1.30 & 0.04 & 8.32   & --111.0 \\
\object{SDSS J233113--010933}  &  83.0480&--57.6812& MS & 2.56 & 0.04 & 8.72   & --101.3 \\ 
\hline

\end{tabular}
\end{table*}

\section{Discussion}

The first thing to remark is the quite good performance 
of the [Fe/H] estimates obtained from the SDSS spectra,
down to about [Fe/H]=$-3.5$. 
Owing to the few objects at lowest metallicity no definite
conclusion can be drawn whether the metallicity estimates based on SDSS
spectra are biased.
However, the low metallicity of \object{SDSS\,J102915+172927} 
\citep{EC_Nature} indicates that this is not very likely.

The fact that at low metallicity the estimate based on SDSS
spectra relies essentially on the \ion{Ca}{ii} K line implies
that the method is prone to underestimate the abundances
of stars that are not enhanced in $\alpha$ elements, as
discussed in \citet{bonifacio11}.

Considering the results of the present paper, 
of \citet{Elisabetta} and \citet{EC_Nature,EC12} together  allows us to
conclude that the metallicities estimated from the SDSS spectra
are reliable, in a statistical sense, at least down to
[Fe/H]=--5. The study of the halo metallicity distribution
based on the SDSS spectra will be adressed in a future paper.

The results on the chemical composition of the
present sample of stars suggests that, by  and large, 
they are ``typical'' halo stars compared
to the sample of \citet{bonifacio09}, as we did in figures
\ref{figalpha} to \ref{fig_sr}. Below we  remark
on the groups of elements.

\subsection{$\alpha$ elements}

In Fig.\,\ref{figalpha} we compare the present results
with those of \citet{bonifacio09}, which were rescaled
to the solar abundances here adopted.
The abundance ratios of our stars show a larger scatter
than those of \citet{bonifacio09} and we interpret this result as 
caused the lower S/N ratio of our spectra.
The [Ca/Fe] ratio in the figure is the straight average 
of the abundance derived from \ion{Ca}{i} lines and the
\ion{Ca}{ii} 370.6024\,nm, when both are available, although
in several cases there is a sizeable difference in the 
two abundances, this could arise from an
incorrect gravity.  The most striking case is that 
of \object{SDSS J144640+124917}, \relax for which  
the \ion{Ca}{i} 445.4779\,nm line provides
[Ca/Fe]=+0.51, while the \ion{Ca}{ii} 370.6024\,nm,
provides [Ca/Fe]=--0.26.
The current version of {\tt MyGIsFOS \relax} adjusts
the gravity only on the iron ionisation equilibrium, use of
other elements is possible in future implementations.
Two stars, \object{SDSS\,J082521+040334} and \object{SDSS\,J154246+054426}
show a negative [Ca/Fe], based only on the 
\ion{Ca}{ii} 370.6024\,nm line, remarkably, they
also show a negative [Ti/Fe], although both have 
standard enhancement of Si and Mg.

\object{SDSS\,J130017+263238}
([Fe/H]=--3.65) shows a value of [Si/Fe], [Mg/Fe] and
[Ca/Fe], which is higher than the mean value of the sample, 
but this is not the case for [Ti/Fe].

The two spectra of \object{SDSS\,J004029+160416} provide very
similar abundances, so that they are not distinguishable in the plot.
Instead the two spectra of \object{SDSS\,J002113-005005} 
imply a difference
of 0.1\,dex in [Fe/H];  to identify that the two points
refer to the same star we connected them with a line
in Fig.\,\ref{figalpha} and all subsequent plots.

\subsection{Even iron peak elements}

The abundances of Ni and Cr behave as expected, 
the decrease of the [Cr/Fe] is probably driven
by NLTE effects, as testified by the two measurements
of \ion{Cr}{ii} lines and discussed in \citet{bonifacio09}
and \citet{ber_ces}.
There is one star, 
\object{SDSS\,J082521+040334}, which has a remarkably high
[Ni/Fe]=+0.48, note that this star also has low [Ca/Fe] and [Ti/Fe].
Three more stars have [Ni/Fe]$\sim +0.3$,
\object{SDSS\,J113528+010848}, \object{SDSS\,J154246+054426} and \object{SDSS\,J230814--085526}.
Considering the errors, all these are compatible with
[Ni/Fe]$\sim 0$ at 2$\sigma$. The straight mean of all  values
is $\langle \rm [Ni/Fe] \rangle = 0.13\pm 0.16$, again compatible with zero.

\subsection{Odd iron peak elements}

 The elements Sc and Co  show a behaviour
that totally agrees with that found by 
\citet{bonifacio09}. Only  Mn is higher on average
and a few stars show a [Mn/Fe] that is almost solar,
while others show [Mn/Fe] as low as $-0.5$\,dex or lower.
Both in the present paper and in \citet{bonifacio09}
the abundances are derived from the \ion{Mn}{i}
resonance triplet at 403\,nm. In the plots we did not adopt 
the correction of +0.4\,dex suggested by \citet{bonifacio09},
based on the offset found in giant stars between resonance
and high excitiation lines.  
\citet{ber_mn} have computed Mn abundances
for a sample of metal--poor dwarfs, and for their
model  closest to the parameters of our stars
(\teff = 6000, \glog = 4.0, [M/H] = -3.0) they found a sizeable
NLTE correction of about 0.6\,dex for the resonance triplet,
but their computations do not show strong variations
of the NLTE corrections with either temperature, metallicity,
or gravity. Although a wider set of computations would be desirable,
it does seem unlikely that the scatter we observe in Mn abundances
is caused by differential NLTE effects.
In Fig.\,\ref{mnplot} we compare
 two stars whose [Mn/Fe] ratios differ by 0.5\,dex,
clearly  the difference of almost 300\,K in effective
temperature between the two stars cannot justify the
obvious difference in line strength.
It is possible that there is some real scatter in Mn abundances, and 
one should keep in mind that also in the sample of
\citet{bonifacio09} there are at least two stars with relatively
high [Mn/Fe], although at slightly higher metallicities. 

\subsection{Strontium }

Strontium is the only neutron capture element 
that we are able to measure in this sample of stars. 
It shows a large scatter in [Sr/Fe] ratios, 
as expected \citep{burris,barklem,patrick},
and is compatible with the results of
\citet{bonifacio09}.

\subsection{Lithium}

We measured lithium abundances for 12 stars and provide four 
upper limits. 
In Figures \ref{limet} and \ref{liteff} we show the lithium abundances in our
stars. 
We also added to the plot the stars 
of \citet{sbordone10} and the two components of the binary system  
\object{CS\,22876-32} \citep[][crosses]{jonay}, the 
three Li-depleted stars \object{G\,122-69},
\object{G\,139-8}, \object{G\,186-26} from \citet[][asterisks]{norris},
star \object{SDSS\,J102915+172927} from 
\citet[][star symbol]{EC_Nature} and star
\object{HE\,1327-2326} from \citet[][crossed square]{Frebel08}.
 
Our measured equivalent width of
the \ion{Li}{i} doublet for star \object{SDSS\,J090733+024608}, 
agrees excellently with that of \citet{Elisabetta}. 
This 
is the only star for which they could measure lithium 
from the X-Shooter spectra.

The present results reinforce the existence
of the meltdown of the Spite plateau at low metallicities,
found by \citet{aoki09} and \citet{sbordone10}. In particular the upper
limits, which are quite stringent, call for strong depletions.
However, in spite of the evident meltdown it is remarkable
that  two of the lowest metallicity stars of the sample, 
\object{SDSS\,J130017+263238}, and
\object{SDSS\,J090733+024608}, 
stay on the Spite plateau, implying that the meltdown is not
a mere increase of Li depletion with decreasing metallicity.

It is intriguing that the two most iron-poor objects in the plot, 
\object{HE\,1327-2326} and  \object{SDSS\,J102915+172927}, display
no lithium. In the Figures \ref{limet}
and \ref{liteff} we plot the upper limits as provided
by \citet{Frebel08} and \citet{EC_Nature}, respectively,
although in the first case it is a $3\sigma$ upper limit, while
in the latter it is a $5\sigma$ upper limit, by adopting the same
criterion the two upper limits would lie roughly at the same level.
\citet{EC_Nature} have pointed out that given the peculiar
chemical composition of  \object{HE\,1327-2326}, there is little
connection between its iron abundance and its metallicity $Z$.
By plotting the lithium abundances as a function of carbon abundance,
rather than of iron abundance,  \object{HE\,1327-2326} occupies
the same region in the diagram as 
the well-known Li-depleted stars
\object{G\,122-69},
\object{G\,139-8} and \object{G\,186-26}.
The known Li-depleted stars are a handful and at 
[Fe/H]$> -3.0$ they are only a few percent of the warm halo
stars. The fact that in a very limited sample, like ours, 
we find three new Li-depleted stars suggests that these
stars are more frequently found at low [Fe/H].
There is no widely accepted explanation of the Li-depleted stars.
Probably the most promising hypothesis is that
of \citet{Ryan02}, who measured sizeable
rotational velocities in some  of the Li-depleted
stars and suggested that they are ``blue stragglers to be'',
i.e. although they have colours that are currently compatible
with the halo turn-off, in the future, as the TO stars evolve to
redder colours, they will remain in their current position in the
colour-magnitude diagram. 
Although it is well known that blue stragglers display no
lithium \citep{glaspey}, the reasons for this, or
for the blue stragglers phenomenon, for that matter,  are
not well understood.
Although promising, the   \citet{Ryan02} hypothesis
does not readily provide any argument that could explain a 
higher frequency of these objects at low [Fe/H].

There are many mechanisms that can destroy lithium, all of which
imply that the material is processed at temperatures exceeding
two million degrees. In the present state of affairs
it is difficult to argue that one and the same mechanism
is responsible for all the Li-depleted stars.
At the same time it is not clear if Li-depleted stars are
related to the Spite plateau meltdown.

Figure\,\ref{liteff} shows that Li-depleted stars are not
confined to a particular range in \teff. The hottest
Li-depleted star has an effective temperature of 6400\,K.

\section{Conclusions}

Thanks to a very efficient technique of selecting
extremely metal poor stars from the Sloan Digital Sky 
Survey, we have
considerably increased the sample of stars at the lowest
metallicities for which detailed chemical abundances
were measured. Note that while in the sample
of \citet{bonifacio09} there was only one dwarf star
with [Fe/H]$<-3.4$, the present sample comprises
six such stars. 
The agreement between the metallicities estimated
from the SDSS spectra and those derived from the
UVES spectra suggests that the former are reliable, 
at least in a statistical sense, for investigating the
metallicity distribution of the halo.

The chemical composition of the sample of stars 
presented here is consistent with that of the sample
of \citet{bonifacio09}, suggesting that the halo
was well mixed at all  probed metallicities.
The only possible exception is a sizeable scatter
in Mn abundances. This indication should be taken with caution however,
 given the lower S/N ratio
of the present data compared to that of \citet{bonifacio09},
and because there are still only  few  observed stars. 

The relatively high S/N ratios in the red part of the spectra
allowed us to increase the sample of dwarf stars with measured
lithium abundance at metallicities below --3.0. 
The meltdown of the Spite plateau is confirmed and the three 
upper limits show the occurance of strong lithium depletions,
at least in some cases. On the other hand, the presence at the
lowest metallicities of non-depleted stars on the Spite plateau
implies that the meltdown depends on some other parameter besides
metallicity.

\begin{acknowledgements}
We acknowledge support from the Programme National
de Physique Stellaire (PNPS) and the Programme National
de Cosmologie et Galaxies (PNCG) of the Institut National de Sciences
de l'Universe of CNRS. HGL acknowledges financial 
support by the Sonderforschungsbereich SFB\,881
``The Milky Way System'' (subproject A4) of the 
German Research Foundation
(DFG).

\end{acknowledgements}

\longtab{1}{
\begin{longtable}{llllllll}
\caption{\label{logobs}Log of the Observations.}\\
\hline\hline
Star           & Program     &Exposure time& central  & slit & Binning &MJD & Airmass\\       
               &             &  time       & wavelength (nm)    & width $''$&& JD-2400000.5\\ 
               &             &   s         & nm             &           & days\\
\hline
\endfirsthead
\caption{continued.}\\
\hline\hline
Star           & Program     &Exposure time& central  & slit & Binning &MJD & Airmass\\       
               &             &  time       & wavelength     & width $''$&& JD-2400000.5\\ 
               &             &   s         & nm             &           && days\\
\hline
\endhead
\hline
\endfoot
SDSS J223143-094834&078.D-0217(A)&3004.999&580&1.4&$2\times 2$&53995.085976&1.107\\
SDSS J223143-094834&078.D-0217(A)&3005.001&390&1.4&$2\times 2$&53995.085987&1.107\\
SDSS J230814-085526&078.D-0217(A)&3600.001&390&1.4&$2\times 2$&53997.256895&1.221\\
SDSS J230814-085526&078.D-0217(A)&3599.999&580&1.4&$2\times 2$&53997.256899&1.221\\
SDSS J223143-094834&078.D-0217(A)&3004.999&580&1.4&$2\times 2$&53998.013442&1.416\\
SDSS J223143-094834&078.D-0217(A)&3005.001&390&1.4&$2\times 2$&53998.013452&1.416\\
SDSS J223143-094834&078.D-0217(A)&3004.999&580&1.4&$2\times 2$&53998.049319&1.203\\
SDSS J223143-094834&078.D-0217(A)&3005.001&390&1.4&$2\times 2$&53998.049330&1.203\\
SDSS J002113-005005&078.D-0217(A)&3239.999&580&1.4&$2\times 2$&53998.165339&1.145\\
SDSS J002113-005005&078.D-0217(A)&3240.001&390&1.4&$2\times 2$&53998.165349&1.145\\
SDSS J002113-005005&078.D-0217(A)&3239.999&580&1.4&$2\times 2$&53998.204093&1.095\\
SDSS J002113-005005&078.D-0217(A)&3240.001&390&1.4&$2\times 2$&53998.204116&1.095\\
SDSS J031745+002304&078.D-0217(A)&3240.001&390&1.4&$2\times 2$&53998.246345&1.306\\
SDSS J031745+002304&078.D-0217(A)&3239.999&580&1.4&$2\times 2$&53998.246361&1.306\\
SDSS J031745+002304&078.D-0217(A)&3239.999&580&1.4&$2\times 2$&53998.285346&1.162\\
SDSS J031745+002304&078.D-0217(A)&3240.001&390&1.4&$2\times 2$&53998.285356&1.162\\
SDSS J002749+140418&078.D-0217(A)&3004.999&580&1.4&$2\times 2$&54002.125063&1.503\\
SDSS J002749+140418&078.D-0217(A)&3005.001&390&1.4&$2\times 2$&54002.125073&1.503\\
SDSS J002749+140418&078.D-0217(A)&3004.999&580&1.4&$2\times 2$&54002.161983&1.342\\
SDSS J002749+140418&078.D-0217(A)&3005.001&390&1.4&$2\times 2$&54002.162032&1.342\\
SDSS J004029+160416&078.D-0217(A)&3004.999&580&1.4&$2\times 2$&54003.185786&1.342\\
SDSS J004029+160416&078.D-0217(A)&3005.001&390&1.4&$2\times 2$&54003.185797&1.342\\
SDSS J004029+160416&078.D-0217(A)&3004.999&580&1.4&$2\times 2$&54003.222459&1.321\\
SDSS J004029+160416&078.D-0217(A)&3005.001&390&1.4&$2\times 2$&54003.222535&1.321\\
SDSS J090733+024608&078.D-0217(A)&3004.999&580&1.4&$2\times 2$&54085.302979&1.159\\
SDSS J090733+024608&078.D-0217(A)&3005.001&390&1.4&$2\times 2$&54085.302990&1.159\\
SDSS J082521+040334&078.D-0217(A)&3136.999&580&1.4&$2\times 2$&54086.305133&1.140\\
SDSS J082521+040334&078.D-0217(A)&3137.001&390&1.4&$2\times 2$&54086.305159&1.140\\
SDSS J082521+040334&078.D-0217(A)&3144.999&580&1.4&$2\times 2$&54100.164555&1.461\\
SDSS J082521+040334&078.D-0217(A)&3145.001&390&1.4&$2\times 2$&54100.164618&1.461\\
SDSS J082521+040334&078.D-0217(A)&3136.999&580&1.4&$2\times 2$&54100.203967&1.250\\
SDSS J082521+040334&078.D-0217(A)&3137.001&390&1.4&$2\times 2$&54100.203982&1.250\\
SDSS J090733+024608&078.D-0217(A)&3004.999&580&1.4&$2\times 2$&54100.284761&1.131\\
SDSS J090733+024608&078.D-0217(A)&3005.001&390&1.4&$2\times 2$&54100.284775&1.131\\
SDSS J113528+010848&078.D-0217(A)&3239.999&580&1.4&$2\times 2$&54132.204936&1.443\\
SDSS J113528+010848&078.D-0217(A)&3240.001&390&1.4&$2\times 2$&54132.204950&1.442\\
SDSS J113528+010848&078.D-0217(A)&3240.000&580&1.4&$2\times 2$&54132.247986&1.216\\
SDSS J113528+010848&078.D-0217(A)&3240.001&390&1.4&$2\times 2$&54132.248000&1.216\\
SDSS J143632+091831&078.D-0217(A)&3004.999&580&1.4&$2\times 2$&54133.350058&1.431\\
SDSS J143632+091831&078.D-0217(A)&3005.001&390&1.4&$2\times 2$&54133.350074&1.431\\
SDSS J144640+124917&078.D-0217(A)&3004.999&580&1.4&$2\times 2$&54158.341745&1.292\\
SDSS J144640+124917&078.D-0217(A)&3005.001&390&1.4&$2\times 2$&54158.341760&1.291\\
SDSS J154246+054426&078.D-0217(A)&3136.999&580&1.4&$2\times 2$&54174.340722&1.181\\
SDSS J154246+054426&078.D-0217(A)&3137.001&390&1.4&$2\times 2$&54174.340741&1.181\\
SDSS J154246+054426&078.D-0217(A)&3136.999&580&1.4&$2\times 2$&54177.235829&1.697\\
SDSS J154246+054426&078.D-0217(A)&3137.001&390&1.4&$2\times 2$&54177.235847&1.697\\
SDSS J154246+054426&078.D-0217(A)&3136.999&580&1.4&$2\times 2$&54177.275076&1.373\\
SDSS J154246+054426&078.D-0217(A)&3137.001&390&1.4&$2\times 2$&54177.275094&1.373\\
SDSS J082118+181931&081.D-0373(A)&3599.997&580&0.6&$1\times 1$&54583.989490&1.435\\
SDSS J082118+181931&081.D-0373(A)&3600.001&390&0.6&$1\times 1$&54583.989571&1.435\\
SDSS J122935+262445&081.D-0373(A)&3599.997&580&0.8&$1\times 1$&54584.036879&1.874\\
SDSS J122935+262445&081.D-0373(A)&3600.001&390&0.8&$1\times 1$&54584.036926&1.874\\
SDSS J130017+263238&081.D-0373(A)&3599.997&580&0.8&$1\times 1$&54584.081867&1.721\\
SDSS J130017+263238&081.D-0373(A)&3600.001&390&0.8&$1\times 1$&54584.081909&1.720\\
SDSS J130017+263238&081.D-0373(A)&3599.997&580&0.8&$1\times 1$&54584.124756&1.596\\
SDSS J130017+263238&081.D-0373(A)&3600.001&390&0.8&$1\times 1$&54584.124810&1.596\\
SDSS J082118+181931&081.D-0373(A)&3599.997&580&0.8&$1\times 1$&54585.007509&1.521\\
SDSS J082118+181931&081.D-0373(A)&3600.001&390&0.8&$1\times 1$&54585.007553&1.521\\
SDSS J153110+095255&081.D-0373(B)&3599.997&580&1.0&$2\times 2$&54700.994156&1.365\\
SDSS J153110+095255&081.D-0373(B)&3600.001&390&1.0&$2\times 2$&54700.994233&1.365\\
SDSS J153110+095255&081.D-0373(B)&4199.997&580&1.0&$2\times 2$&54701.036850&1.669\\
SDSS J153110+095255&081.D-0373(B)&4200.001&390&1.0&$2\times 2$&54701.036954&1.669\\
SDSS J233113-010933&081.D-0373(B)&3599.997&580&1.0&$2\times 2$&54701.089799&2.109\\
SDSS J233113-010933&081.D-0373(B)&3600.001&390&1.0&$2\times 2$&54701.089846&2.109\\
SDSS J233113-010933&081.D-0373(B)&3599.997&580&1.0&$2\times 2$&54701.132475&1.506\\
SDSS J233113-010933&081.D-0373(B)&3600.002&390&1.0&$2\times 2$&54701.132554&1.506\\
SDSS J233113-010933&081.D-0373(B)&3599.998&580&1.0&$2\times 2$&54701.175114&1.239\\
SDSS J233113-010933&081.D-0373(B)&3600.002&390&1.0&$2\times 2$&54701.175154&1.239\\
SDSS J002113-005005&081.D-0373(B)&3599.997&580&0.7&$1\times 1$&54701.219980&1.203\\
SDSS J002113-005005&081.D-0373(B)&3600.001&390&0.7&$1\times 1$&54701.220019&1.203\\
SDSS J004029+160416&081.D-0373(B)&3599.997&580&1.0&$2\times 2$&54701.264875&1.361\\
SDSS J004029+160416&081.D-0373(B)&3600.002&390&1.0&$2\times 2$&54701.264915&1.361\\
SDSS J002113-005005&081.D-0373(B)&3599.997&580&0.7&$1\times 1$&54701.343975&1.162\\
SDSS J002113-005005&081.D-0373(B)&3600.001&390&0.7&$1\times 1$&54701.344085&1.162\\
SDSS J002113-005005&081.D-0373(B)&2623.997&580&0.7&$1\times 1$&54701.386599&1.337\\
SDSS J002113-005005&081.D-0373(B)&2620.002&390&0.7&$1\times 1$&54701.386639&1.337\\
\hline
\end{longtable}
}


\begin{thebibliography}{}

\bibitem[Adelman-McCarthy et al.(2007)]{dr5} 
Adelman-McCarthy, J.~K., et al.\ 2007, \apjs, 172, 634 


\bibitem[Adelman-McCarthy et al.(2008)]{dr6}
Adelman-McCarthy, J.~K., et al.\ 2008, \apjs, 175, 297


\bibitem[Aoki et al.(2009)]{aoki09} Aoki, W., Barklem, P.~S., 
Beers, T.~C., et al.\ 2009, \apj, 698, 1803 

\bibitem[Barklem et 
al.(2005)]{barklem} Barklem, P.~S., Christlieb, N., Beers, T.~C., et al.\ 2005, \aap, 439, 129 


\bibitem[Behara et 
al.(2010)]{Behara} Behara, N.~T., Bonifacio, P., Ludwig, H.-G., 
Sbordone, L., Gonz{\'a}lez Hern{\'a}ndez, J.~I., \& Caffau, E.\ 2010, \aap, 513, A72 



\bibitem[Belokurov et al.(2006)]{belokurov} Belokurov, V., et 
al.\ 2006, \apjl, 642, L137 


\bibitem[Bergemann 
\& Cescutti(2010)]{ber_ces} Bergemann, M., \& Cescutti, G.\ 2010, \aap, 522, A9 

\bibitem[Bergemann 
\& Gehren(2008)]{ber_mn} Bergemann, M., \& Gehren, T.\ 2008, \aap, 492, 823 



\bibitem[Bonifacio \& Caffau(2003)]{BC03} Bonifacio, P., \& Caffau, E.\ 
2003, \aap, 399, 1183 

\bibitem[Bonifacio et 
al.(2002)]{bonifacio02} Bonifacio, P., et al.\ 2002, \aap, 390, 91 


\bibitem[Bonifacio et 
al.(2009)]{bonifacio09} Bonifacio, P., et al.\ 2009, \aap, 501, 519 

\bibitem[Bonifacio et al.(2010)]{bonifacio_rio} Bonifacio, P., 
Caffau E., Ludwig, H.-G., Sbordone, L., Gonz\'alez~Hern\'andez, J.I.,
Behara, N. T., 2009, IAU XVII General Assembly, Joint Discussion 5, ed.
J. Binney

\bibitem[Bonifacio et al.(2011)]{bonifacio11} Bonifacio, P., et 
al.\ 2011, Astronomische Nachrichten, 332, 251 

\bibitem[Bromm 
\& Loeb(2003)]{BL} Bromm, V., \& Loeb, A.\ 2003, \nat, 425, 812 


\bibitem[Burris et al.(2000)]{burris} Burris, D.~L., 
Pilachowski, C.~A., Armandroff, T.~E., et al.\ 2000, \apj, 544, 302 


\bibitem[Caffau et al.(2011a)]{caffau_sol} Caffau, E., Ludwig, 
H.-G., Steffen, M., Freytag, B., \& Bonifacio, P.\ 2011a, \solphys, 268, 255 


\bibitem[Caffau et 
al.(2011b)]{Elisabetta} Caffau, E., Bonifacio, P., Fran{\c c}ois, P., et al.\ 2011b, \aap, 534, A4 



\bibitem[Caffau et al.(2011c)]{EC_Nature} Caffau, E., Bonifacio, 
P., Fran{\c c}ois, P., et al.\ 2011c, \nat, 477, 67 

\bibitem[Caffau et al.(2012)]{EC12} Caffau, E., Bonifacio, 
P., Fran{\c c}ois, P., et al.\ 2012, A\&A in press,  arXiv:1203.2607 


\bibitem[Carollo et al.(2007)]{Carollo07} Carollo, D., et al.\ 
2007, \nat, 450, 1020 

\bibitem[Carollo et al.(2010)]{Carollo10} Carollo, D., et al.\ 
2010, \apj, 712, 692 

\bibitem[Castelli(2005)]{Castelli05} Castelli, F.\ 2005, Memorie 
della Societ\`a Astronomica Italiana Supplementi, 8, 25 

\bibitem[Cayrel(1988)]{cayrel88} Cayrel, R.\ 1988, in 
IAU Symp. 132 The Impact of  Very High S/N
Spectroscopy on Stellar Physics, G. Cayrel de Strobel and M. Spite eds.,
p. 345 


\bibitem[Dekker et al.(2000)]{uves} Dekker, H., D'Odorico, 
S., Kaufer, A., Delabre, B., \& Kotzlowski, H.\ 2000, \procspie, 4008, 534 


\bibitem[D'Odorico et al.(2006)]{X-Shooter} D'Odorico, S., et 
al.\ 2006, \procspie, 6269,  



\bibitem[Eggen et al.(1962)]{ELS} Eggen, O.~J., 
Lynden-Bell, D., \& Sandage, A.~R.\ 1962, \apj, 136, 748 

\bibitem[Font et al.(2011)]{font} Font, A.~S., McCarthy, 
I.~G., Crain, R.~A., Theuns, T., Schaye, J., Wiersma, R.~P.~C., 
\& Dalla Vecchia, C.\ 2011, arXiv:1102.2526 

\bibitem[Fran{\c c}ois et 
al.(2007)]{patrick} Fran{\c c}ois, P., Depagne, E., Hill, V., et al.\ 2007, \aap, 476, 935 



\bibitem[Frebel et al.(2007)]{Frebel} Frebel, A., Johnson, 
J.~L., \& Bromm, V.\ 2007, \mnras, 380, L40 

\bibitem[Frebel et al.(2008)]{Frebel08} Frebel, A., Collet, R., 
Eriksson, K., Christlieb, N., \& Aoki, W.\ 2008, \apj, 684, 588 





\bibitem[Freytag et al.(2002)]{freytag02} Freytag, B., Steffen, 
M., \& Dorch, B.\ 2002, Astronomische Nachrichten, 323, 213


\bibitem[Freytag et al.(2012)]{freytag11} Freytag, B., Steffen, 
M., Ludwig, H.-G., et al.\ 2012, Journal of Computational Physics, 231, 919 

\bibitem[Girardi et 
al.(2002)]{girardi} Girardi, L., Bertelli, G., Bressan, A., et al.\ 2002, \aap, 391, 195 

\bibitem[Glaspey et al.(1994)]{glaspey} Glaspey, J.~W., 
Pritchet, C.~J., \& Stetson, P.~B.\ 1994, \aj, 108, 271 





\bibitem[Gonz{\'a}lez Hern{\'a}ndez et 
al.(2008)]{jonay} Gonz{\'a}lez Hern{\'a}ndez, J.~I., et al.\ 2008, \aap, 480, 233 




\bibitem[Helmi(2008)]{helmi2008} Helmi, A.\ 2008, \aapr, 15, 145 

\bibitem[{{Kurucz}(1993b)}]{K93}
{Kurucz}, R. 1993, SYNTHE Spectrum Synthesis Programs and Line
Data.~Kurucz CD-ROM No.~18.~Cambridge, Mass.: Smithsonian Astrophysical
Observatory, 1993., 18

\bibitem[{{Kurucz}(2005)}]{K05}
{Kurucz}, R.~L. 2005, Memorie della Societ\`a Astronomica
  Italiana Supplementi, 8, 14



\bibitem[Li et 
al.(2010)]{HES_MDF2} Li, H.~N., et al.\ 2010, \aap, 521, A10 



\bibitem[Lodders et al.(2009)]{lodders} Lodders, K., Palme, H., 
\& Gail, H.-P.\ 2009, Landolt-B{\"o}rnstein - 
Group VI Astronomy and Astrophysics Numerical 
Data and Functional Relationships in Science and 
Technology Volume 4B: Solar System.~ ed. 
J.E.~Tr{\"u}mper, 44 



\bibitem[Ludwig et al.(2008)]{Ludwig08} Ludwig, H.-G., 
Bonifacio, P., Caffau, E., Behara, N.~T., Gonz{\'a}lez Hern{\'a}ndez, 
J.~I., \& Sbordone, L.\ 2008, Physica Scripta Volume T, 133, 014037 

\bibitem[Ludwig et al.(2009)]{Ludwig09} Ludwig, H.-G., Caffau, 
E., Steffen, M., Freytag, B., Bonifacio, P., 
\& Ku{\v c}inskas, A.\ 2009, \memsai, 80, 711 





\bibitem[Majewski(1992)]{Majewski} Majewski, S.~R.\ 1992, \apjs, 78, 87 

\bibitem[Marigo et al. (2008)]{marigo} Marigo, P., et al.\ 2008, \aap, 482, 88

\bibitem[Mashonkina et al.(2007)]{mashonkina} Mashonkina, L., 
Korn, A.~J., \& Przybilla, N.\ 2007, \aap, 461, 261 

\bibitem[Norris et al.(1997)]{norris} Norris, J.~E., Ryan, 
S.~G., Beers, T.~C., \& Deliyannis, C.~P.\ 1997, \apj, 485, 370 


\bibitem[Piau et al.(2006)]{piau} Piau, L., Beers, T.~C., 
Balsara, D.~S., Sivarani, T., Truran, J.~W., 
\& Ferguson, J.~W.\ 2006, \apj, 653, 300 

\bibitem[Richards et al.(2002)]{Richards} Richards, G.~T., et 
al.\ 2002, \aj, 123, 2945 



\bibitem[Ryan 
\& Norris(1991)]{RN91} Ryan, S.~G., \& Norris, J.~E.\ 1991, \aj, 101, 1865 

\bibitem[Ryan et al.(2002)]{Ryan02} Ryan, S.~G., Gregory, 
S.~G., Kolb, U., Beers, T.~C., \& Kajino, T.\ 2002, \apj, 571, 501 




\bibitem[Salvadori et al.(2007)]{Salvadori} Salvadori, S., 
Schneider, R., \& Ferrara, A.\ 2007, \mnras, 381, 647 

\bibitem[{{Sbordone}(2005)}]{SB05}
{Sbordone}, L. 2005, Memorie della Societ\`a Astronomica Italiana Supplementi, 8, 61


\bibitem[{{Sbordone} {et~al.}(2004){Sbordone}, {Bonifacio}, {Castelli}, \&
  {Kurucz}}]{SB04}
{Sbordone}, L., {Bonifacio}, P., {Castelli}, F., \& {Kurucz}, R.~L. 2004,
  Memorie della Societ\`a Astronomica Italiana Supplementi, 5, 93


\bibitem[Sbordone et 
al.(2010a)]{sbordone10} Sbordone, L., et al.\ 2010a, \aap, 522, A26 


\bibitem[Sbordone et al.(2010b)]{sbordone_nic} Sbordone, L., 
Bonifacio, P., Caffau, E., \& Ludwig, H.-G.\ 2010, arXiv:1009.5210 

\bibitem[Sbordone et al.(2011)]{mygisfos} Sbordone, L. eta al.\
in preparation

\bibitem[Schneider et al.(2003)]{Schneider} Schneider, R., 
Ferrara, A., Salvaterra, R., Omukai, K., \& Bromm, V.\ 2003, \nat, 422, 869 



\bibitem[Sch{\"o}rck et 
al.(2009)]{HES_MDF1} Sch{\"o}rck, T., et al.\ 2009, \aap, 507, 817 


\bibitem[Searle 
\& Zinn(1978)]{SZ78} Searle, L., \& Zinn, R.\ 1978, \apj, 225, 357


\bibitem[Wedemeyer et 
al.(2004)]{wedemeyer04} Wedemeyer, S., Freytag, B., Steffen, M., Ludwig, H.-G., \& Holweger, H.\ 2004, \aap, 414, 1121 

\bibitem[Yanny et al.(2009)]{segue} Yanny, B., Rockosi, C., 
Newberg, H.~J., et al.\ 2009, \aj, 137, 4377 


\bibitem[York et al.(2000)]{sdss} York, D.~G., et al.\ 2000, 
\aj, 120, 1579 




\bibitem[Zolotov et al.(2009)]{zolotov} Zolotov, A., Willman, 
B., Brooks, A.~M., Governato, F., Brook, C.~B., Hogg, D.~W., Quinn, T., 
\& Stinson, G.\ 2009, \apj, 702, 1058 



\end{thebibliography}
 \end{document}